\documentclass[fleqn,usenatbib]{mnras}

\usepackage{newtxtext,newtxmath}

\usepackage[T1]{fontenc}

\DeclareRobustCommand{\VAN}[3]{#2}
\let\VANthebibliography\thebibliography
\def\thebibliography{\DeclareRobustCommand{\VAN}[3]{##3}\VANthebibliography}

\usepackage[mathscr]{euscript}
\usepackage{graphicx}	
\usepackage{amsmath}
\usepackage[dvipsnames]{xcolor}
\usepackage[nice]{nicefrac}
\newcommand{\orcid}[1]{\href{https://orcid.org/#1}{\includegraphics[width=8pt]{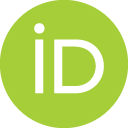}}}

\title[GWs from ZLK triples]{Gravitational-Wave Signatures of Highly Eccentric Stellar-Mass Binary Black Holes in Galactic Nuclei }

\author[Grishin et al.]{
Evgeni Grishin,$^{1,2}$ 
\orcid{0000-0001-7113-723X}%
\thanks{E-mail: evgeni.grishin@monash.edu}
Isobel M. Romero-Shaw,$^{3,4,5,6}$
\thanks{E-mail:romero-shawi@cardiff.ac.uk}
\orcid{0000-0002-4181-8090}
and Alessandro A. Trani,$^{7,8}$
\orcid{0000-0001-5371-3432}
\\
$^{1}$School of Physics and Astronomy, Monash University, Clayton, VIC 3800, Australia\\
$^{2}$OzGrav: Australian Research Council Centre of Excellence for Gravitational Wave Discovery, Clayton, VIC 3800, Australia\\
$^3$Department of Applied Mathematics and Theoretical Physics, Cambridge CB3 0WA, United Kingdom\\
$^4$Kavli Institute for Cosmology Cambridge, Madingley Road Cambridge CB3 0HA, United Kingdom\\
$^5$H. H. Wills Physics Laboratory, University of Bristol, Tyndall Avenue, Bristol BS8 1TL, UK\\
$^6$Gravity Exploration Institute, School of Physics and Astronomy, Cardiff University, Cardiff, CF24 3AA, UK\\
$^7$Departamento de Astronom\'ia, Facultad Ciencias F\'isicas y Matem\'aticas, Universidad de Concepci\'on, Concepci\'on, 4030000, Chile \\
$^8$National Institute for Nuclear Physics – INFN, Sezione di Trieste, I-34127, Trieste, Italy \\
}

\date{Accepted XXX. Received YYY; in original form ZZZ}

\pubyear{2024}

\begin{document}
\label{firstpage}
\pagerange{\pageref{firstpage}--\pageref{lastpage}}
\maketitle

\begin{abstract}
A significant fraction of compact-object mergers in galactic nuclei are expected to be eccentric in the Laser Interferometer Space Antenna (LISA) frequency sensitivity range, $10^{-4} - 10^{-1}\ \rm Hz$. Several compact binaries detected by the LIGO-Virgo-KAGRA Collaboration may retain hints of residual eccentricity at $\sim 10$ Hz, suggesting dynamical or triple origins for a significant fraction of the gravitational-wave-observable population. In triple systems, von-Zeipel-Lidov-Kozai oscillations perturb both the eccentricity and the argument of pericentre, $\omega$, of the inner black hole binary. The latter could be fully \textit{circulating}, where $\omega$ cycles through $2\pi$, or may \textit{librate}, with $\omega$ ranges about a fixed value with small or large variation. We use \texttt{TSUNAMI}, a regularised N-body code with up to 3.5 post-Newtonian (PN) term corrections, to identify four different families of orbits: (i) circulating, (ii) small and (iii) large amplitude librating, and (iv) merging orbits. We develop and demonstrate a new method to construct gravitational waveforms using the quadrupole formula utilising the instantaneous {\it total} acceleration of each binary component in \texttt{TSUNAMI}. We show that the four orbital families have distinct waveform phenomenologies, enabling them to be distinguished if observed in LISA. The orbits are also distinguishable from an isolated binary or from a binary perturbed by a different tertiary orbit, even if the secular timescale is the same. Future burst timing models will be able to distinguish the different orbital configurations. For efficient binary formation, about $\sim 1000$ binaries can have highly eccentric, librating orbits in the Galactic Centre. 

\end{abstract}




\section{Introduction}

Since the discovery of GW150914 \citep{GW150914}, detections of binary compact objects via their gravitational-wave (GW) emission have revolutionised modern-day astronomy. Over two hundred stellar-mass binary black hole (BBH) mergers have been detected and announced by the LIGO-Virgo-KAGRA (LVK) collaboration so far \citep{GWTC-4}, with several other collaborations finding evidence for additional mergers in the data \citep{zackay1, zackay2, zackay3}.
While the LVK detectors are sensitive to the $\sim10\ \rm Hz$--$2000\ \rm Hz$ band and capture the final few seconds of a BBH inspiral and merger, the sensitivity of the Laser Interferometer Space Antenna (LISA) is in the range $\sim10^{-4}\ \rm Hz$ to $\sim10^{-1}\ \rm Hz$. Planned to launch in 2035, LISA will detect GWs from stellar-mass (and higher-mass) compact binaries years to decades before they merge \citep{Klein:2022:lastthreeyears}, along with other GW events such as extreme mass-ratio inspirals (see \citealp{LISA2023} and references therein).  

The astrophysical origins of gravitational-wave (GW) mergers are generally divided into two broad categories: \textit{isolated} \citep{belczynski16, neijssel19} and \textit{dynamical}. In the isolated channel, binaries evolve without significant external perturbations throughout their lifetimes. This scenario is expected to produce mergers with small, aligned spins and nearly circular orbits at $10$~Hz GW frequency, and with component masses limited by pair instability \citep{heger1, heger2}.

In contrast, dynamical formation occurs through gravitational interactions in densely-populated stellar environments. These interactions can assemble binaries with eccentric orbits, large or misaligned spins, and component masses exceeding the pair-instability limit through hierarchical mergers \citep[see][for details]{mandel22}. Dynamical channels include a range of environments, such as open clusters \citep{ku19,trani2021}, globular clusters \citep{rod16, sam18}, and nuclear star clusters \citep{antonini12, gri18, hoang2018}. Black holes embedded in the dense accretion discs of active galactic nuclei (AGN) may also merge at significant rates \citep{stone17, tagawa2020, samsing_nat,trani2024a}, although the relevant physical processes remain uncertain \citep{gilbaum2022, ggs24, gilbaum2025, moncrieff25}.

Recent analyses of the masses and/or spins of observed stellar-mass BBHs suggest that some fraction of the observed population are hierarchical in origin, containing one or more remnants of previous mergers \citep{ant24, Tong2025, Antonini2025}; such mergers may only occur in densely-populated environments. Several BBH events also show tentative evidence for measurable BBH eccentricity at detection \citep[e.g.,][]{ecc3, gupte24, 2025arXiv250415833D}. However, confident measurements of eccentricity remain elusive, given the lack of complete inspiral–merger–ringdown waveform models incorporating both the effects of spin-induced precession and eccentricity \citep{ecc4, 2024PhRvD.109d3037D}. Detecting orbital eccentricity in LVK detectors is considered ``smoking gun'' evidence that the binary did not evolve in complete isolation \citep[e.g.][]{rod18, f19, samsing_nat}: it may have evolved in one of the dynamical environments previously introduced, or in a field triple.


Hierarchical field triples, in which a distant tertiary perturbs the inner binary \citep{antonini17, ll19,  gp22, ginat2025}, represent a middle ground between the isolated and dynamical evolution channels for BBH mergers. These systems are generally long-lived; however, their evolution is influenced by complex dynamics, similar to dynamically-formed binaries. The inner binary merger may be accelerated by mass transfer and encouraged by common-envelope or chemically homogeneous evolution, important evolutionary mechanisms in the isolated channel that facilitate observable BBH mergers \citep{dor24, kummer24}.  The merger time may \textit{also} be accelerated through triple dynamics, which can heighten eccentricity through von-Zeipel-Lidov-Kozai (ZLK) oscillations \citep[e.g.,][]{vg25}. GW merger followed by a natal kick and stellar merger could lead to EM counterparts \citep{naoz2025}. Therefore, mergers that form through isolated triples bear characteristics of both isolated and dynamical binaries: limited masses, but a wider distribution of spins than isolated binaries, and the potential for detectable eccentricity at $10$~Hz \citep[e.g.,][]{RodriguezAntonini:2018:triples,trani2022, Martinez:2022:massratio, 2025arXiv250723212D, steg25}.

Theoretical modelling suggests that a fraction of events from various dynamical channels retain non-negligible eccentricity at detection in LVK instruments \citep[e.g.][]{rod18, f19, samsing_nat}. If this fraction is small, as it is predicted to be at least in the case of dense clusters ($\sim5\%$) and field triples ($\sim10\%$) \citep[see][and references therein]{2025arXiv250723212D} it may take LVK detectors a long time to observe a number of eccentric events that allows us to map cleanly from observed BBH eccentricity distribution to formation mechanism. \citet{ecc3} predicts $\gtrsim$ 80 eccentric events are needed to identify a dominant eccentric merger formation mechanism; currently there are tentative claims of $\lesssim$ 5 \citep[e.g.,][]{ecc3, gupte24, 2025arXiv250415833D}.

Meanwhile, going to lower, LISA-like frequencies allows these populations to be separated cleanly. 
While dynamically-formed mergers that retain detectable eccentricity at $10$~Hz will either become bound at higher frequencies than LISA can access or strictly lose eccentricity via GW emission while in the LISA band \citep{2018MNRAS.481.4775D}, tertiary-driven BBHs undergoing ZLK oscillations may oscillate from high to low frequencies for years in LISA \citep[e.g.,][]{hoang19, randall19, knee2024}, leading to distinctly different signal morphologies. These signals from triples can be further characterised into different families, which we explore in this paper.

For galactic nuclei, the perturbation of stellar BHs by the supermassive BH (SMBH) induces periodic ZLK oscillations of the eccentricity $e_1$ of the inner orbit and the mutual inclination $\iota_{\rm mut}$ between the two orbits \citep{vZ1910,lid62,koz62}. The corresponding angle of the oscillation is the inner argument of pericenter $\omega$, which could either librate around an equilibrium value or complete a full circulation (e.g. \citealp{naoz13}). A hallmark of ZLK oscillations is that the peak eccentricity depends only on the initial inclination via $e_{\rm max} = \sqrt{1 - 5 \cos^2 \iota_{\rm mut}/3}$ for an initially almost circular orbit. However, this is true only when $\omega$ is circulating, while a librating solution will have somewhat limited eccentricity oscillations (e.g. \citealp{gri2024a}). 

The gravitational-wave (GW) signatures of a BBH in the Galactic Centre being perturbed by Sagittarius A* were recently studied by \cite{knee2024}, where the secular approximation was utilised for an individual orbit, followed by a wavelet-like transform for the burst-like time evolution of the strain polarisations. However, several orbital families are possible, including orbits with small libration around a fixed point in $e_1$-$\omega$ space. Moreover, the secular approximation often breaks down for highly eccentric orbits, where the GW strain frequency is maximal.

In this paper, we identify different orbital families undergoing ZLK oscillations, which have unique characteristics in their dynamical evolution. We use direct N-body integrations using the regularised N-body code \texttt{TSUNAMI} \citep{tsunami-code} with 3.5 post-Newtonian corrections, and calculate the GW strains using the direct acceleration from the quadrupole formula \citep{maggiore08}. We find that the strain amplitudes are generally bursty, and advocate for developing minimally-modelled burst timing searches in the future.

This paper is organised as follows: In sec. \ref{sec2} we review the conditions and dynamics of binaries in galactic nuclei and, in particular, near Sgr A*. In sec. \ref{sec3} we identify the different orbital families. In sec. \ref{sec4} we discuss the observability and astrophysical implications. Finally, in sec. \ref{sec5} we summarise our main finding and discuss future prospects. We find that $\sim 50$ BBH can be librating in the Galactic Centre; the GW signal is generally bursty, due to the highly eccentric orbits; the secular approximation is insufficient to accurately describe the waveforms; and that different orbital configurations have waveform phenomenologies and characteristics that are likely to be distinguishable from each other in LISA observations.

\section{Binary Dynamics in Galactic Nuclei} \label{sec2}

Here we  discuss the motivation for our choice of parameters and overview the general triple dynamics. 

\subsection{Initial conditions}

\subsubsection{Location of the outer binary orbit}


Observations of the Galactic Centre reveal multiple stellar populations: a warped, clockwise disk of young, massive O/WR stars on moderately eccentric orbits \citep[e.g.][]{pau06,bar09,bar10,yelda2014,fellenberg2022} and an isotropic population of B-type (called S-stars) on high-eccentricity orbits in the central ${\sim}0.04$~pc \citep[e.g.][]{levin03,lu13,do13}. Many of the young stars may reside in binaries \citep{pfu14,naoz18}. For the clockwise disk specifically, the leading interpretation is in-situ star formation following the capture and circularisation of an infalling gas cloud and fragmentation of a transient, dense accretion disk \citep[e.g.][]{mapelli2016,trani2018}. By contrast, the S-star population likely requires subsequent dynamical processes (e.g. binary disruption, resonant relaxation, secular torques) to explain their orbital properties \citep[e.g.][]{per09,trani2019,perna22}. 

The sphere of influence of Sgr A* could contain up to $2\times10^4$ BHs \citep{me00, freitag06, hailey2018, rose22}. 
The general distribution of stars near Sgr A* is expected to follow a power-law Bahcall-Wolf cusp of $\rho \propto r^{-7/4}$ \citep{bahcall76,bahcall77}. The closest luminous stars are located in the S-star cluster around $a \gtrsim 0.12^{\prime\prime} \approx 10^3\ \rm au$ \citep{fellenberg2022}. On the other hand, the heavier BH population is expected to be more centrally concentrated due to mass segregation in nuclear star clusters \citep[e.g.][]{alexander09, zhang24}, thus BH can reside in an inner cluster with $a_{\rm BH} \lesssim 10^3 \ \rm au$. Recent analysis of the Gravity collaboration has excluded IMBH with $M \lesssim 2\times10^3 M_\odot$ within the orbit of S2 \citep{gravity_imbh_2023}, thus the BH population is expected to be in the form of stellar mass BHs. 


Active galactic nuclei (AGN) with significant luminosity possess massive accretion discs which extend to about $10^5R_{\rm g}$, where $R_{\rm g} = GM_{\rm SMBH}/c^2 = 0.04 (M/4\times 10^6 M_\odot)\ \rm au$ is the gravitational radius. BHs can form either in-situ from massive star evolution after supernova explosion within AGN discs \citep{sne_agn2, gri21}, or be captured by the AGN discs after the BH forms \citep{aat2025}. BHs tend to accumulate at migration traps, which are located around $\lesssim 10^3 R_g$ \citep{bellovary16}. However, \cite{ggs24} has recently revised the trap locations to be between $\sim 700R_{\rm g}$ and $10^5 R_{\rm g}$, depending on the mass and accretion rate of the SMBH.

\textbf{To summarise}, stellar-mass BHs tend to lie at a range of $10^3-10^4R_{\rm g}$ for both quiescent and active galactic nuclei. We focus on the range of $50-100\ \rm au$ for a Sgr A*-like SMBH of mass $4\times 10^6 M_\odot$ (which corresponds to $(1.25-2.5) \times 10^3 R_{\rm g}$).

\subsubsection{Inner binary orbits}

\begin{figure}
\includegraphics[width=0.49\textwidth]{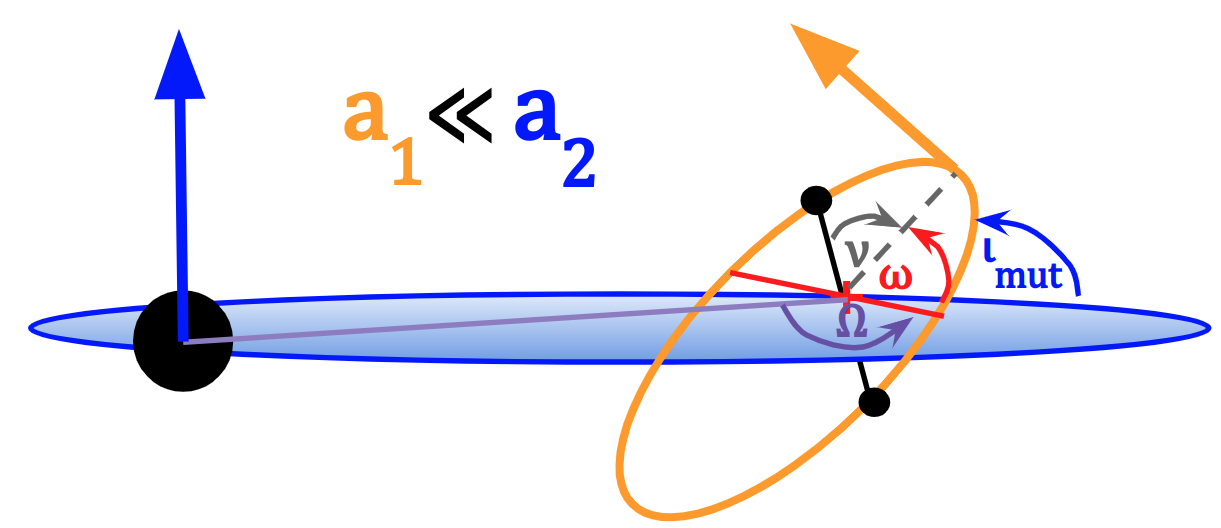}
    \caption{Sketch of the hierarchical triple system (not to scale). The inner binary orbit (orange) is inclined with inclination $\iota_{\rm mut}$ to the SMBH orbital plane. The argument of pericentre $\omega$ is measured from the line of nodes (red) to the pericentre in the inner binary plane. The nodal angle $\Omega$ is measured from the tertiary SMBH to the line of nodes in the outer binary plane.}\label{sketch}
\end{figure}

Consider a binary of masses $m_1$ and $m_2$ (or total mass $m_{\rm bin} = m_1 + m_2$) with semi-major axis $a_1$ and initial eccentricity $e_1$. The centre of mass revolves around a central SMBH of mass $m_3 \gg m_1+m_2$ on semi-major axis $a_2$ and eccentricity $e_2$. The mutual inclination between the two orbital planes is $\iota_{\rm mut}$. Figure \ref{sketch} sketches the orbital elements of the hierarchical triple system. If the tertiary is too close, the binary is dynamically unstable. Many studies have been devoted to stability of triple systems \citep[e.g.,][]{mar01, gri17, he18, tory22, lalande2022, vyn22, hayashi22, hayashi23}. In our case, the measure of stability is encapsulated in the Hill radius,

\begin{equation}
    r_{\rm H} = a_2(1-e_2) \left(\frac{m_{\rm bin}}{3m_3} \right)^{1/3} = 0.9 \bar{a}_2 \bar{m}_{\rm bin}^{1/3} \bar{m}_3^{-1/3}\ \rm au , \label{eq:rhill}
\end{equation} 
where $e_2$ is assumed to be $0$ and the other values are normalised to $\bar{a}_2 = a_2/60\ \rm au$, $\bar{m}_{\rm bin} = m_{\rm bin} / 40M_\odot$, and $\bar{m}_3 = 4\times 10^6 M_\odot$.  

Relativistic corrections induce extra precession on the argument of pericentre $\omega$, which may quench effects of the tertiary SMBH. The relative strength of GR precession is encapsulated in 
\begin{equation}
    \epsilon_{\rm GR} \equiv \left|  \frac{\dot{\omega}_{\rm GR}}{\dot{\omega}_{\rm ZLK}} \right|  = \frac{3m_{\rm bin}}{m_3}\bigg(\frac{a_2}{a_1}\bigg)^3\frac{r_g}{a} = 0.005 \times \bar{a}_1^{-4} \bar{m}_{\rm bin}^2 \bar{m}_3^{-1} \bar{a}_2^3, \label{eps_gr}
\end{equation}
where $\bar{a}_1= a_1/0.15\ \rm au$ and $r_g = G m_{\rm bin}/c^2$ is the mutual gravitational radius. Generally, systems with $\epsilon_{\rm GR} \ge 1$ (or $a_1 \le 0.04\ \rm au$) have tertiary effects quenched, and evolve as isolated binaries, effectively decoupled from the influence of the tertiary. Meanwhile $\epsilon_{\rm GR} \ll 1$ could exhibit large eccentricity variations. The actual maximal eccentricity in the presence of GR corrections has been derived in several studies \citep[e.g.,][]{gri18, man22}.

\textbf{To summarise}, the range of inner binary orbits where the SMBH is able to perturb the orbit but still keep it stable spans around one order of magnitude, $0.04\ \rm au \lesssim a_1\rm\ \lesssim 0.4\ \rm au$. The upper limit is due to the Hill stability $a_1 \lesssim 0.5r_{\rm H}$ (Eq. \ref{eq:rhill}), while the lower limit is due to transition to 1PN-GR dominated precession (Eq. \ref{eps_gr}), where the binary will effectively evolve in isolation.


\subsection{Properties of highly inclined triple systems}

For a dynamically stable binary dominated by the SMBH over GR ($\epsilon_{\rm GR} \le 1)$, quasi-periodic ZLK oscillations occur if the mutual inclination $\iota_{\rm mut}$ is large enough. For almost co-planar orbits, $\omega$ is always circulating and completes a full revolution in a secular timescale. Beyond a critical inclination, a bifurcation in the phase portrait leads to a fixed point in $e_1$-$\omega$ space and orbits can either circulate (where $\omega$ completes a full revolution) or librate (where $\omega$ is limited in range, see e.g., \citealp{naoz16}). This is somewhat analogous to the pendulum model 
\citep[see][for a detailed analysis]{basha_zlk}\footnote{The 'angle' in \cite{basha_zlk} is not $\omega$, but defined via a combination of the eccentricity and angular momentum vectors (see their Eq. 25). The 'momentum' is proportional to the $z$ component of the eccentricity vector, while the potential energy (i.e. the 'height' of the pendulum) is related to the eccentricity $e_1$. Moreover, circulation of $\omega$ corresponds to a libration in the pendulum model, and vice versa (see their Figure 1 and 2).}. There is a critical curve that lies on the unstable fixed point (where the pendulum is balanced on the top) that separates between circulating and librating orbits, which is called the \textit{separatrix}. \cite{hansen20} studied the structure of fixed and saddle points of the hierarchical three body problem to octupole order where the additional dependence on the outer eccentricity and the nodal angle lead to a diverse orrery of families of fixed and saddle points. Here we focus on circular outer orbit with equal mass inner binaries, thus the octupole term is absent, but we take into account the corrections due to mild hierarchies, as explained below:
\begin{figure*}
    \centering
    \includegraphics[width=0.98\textwidth]{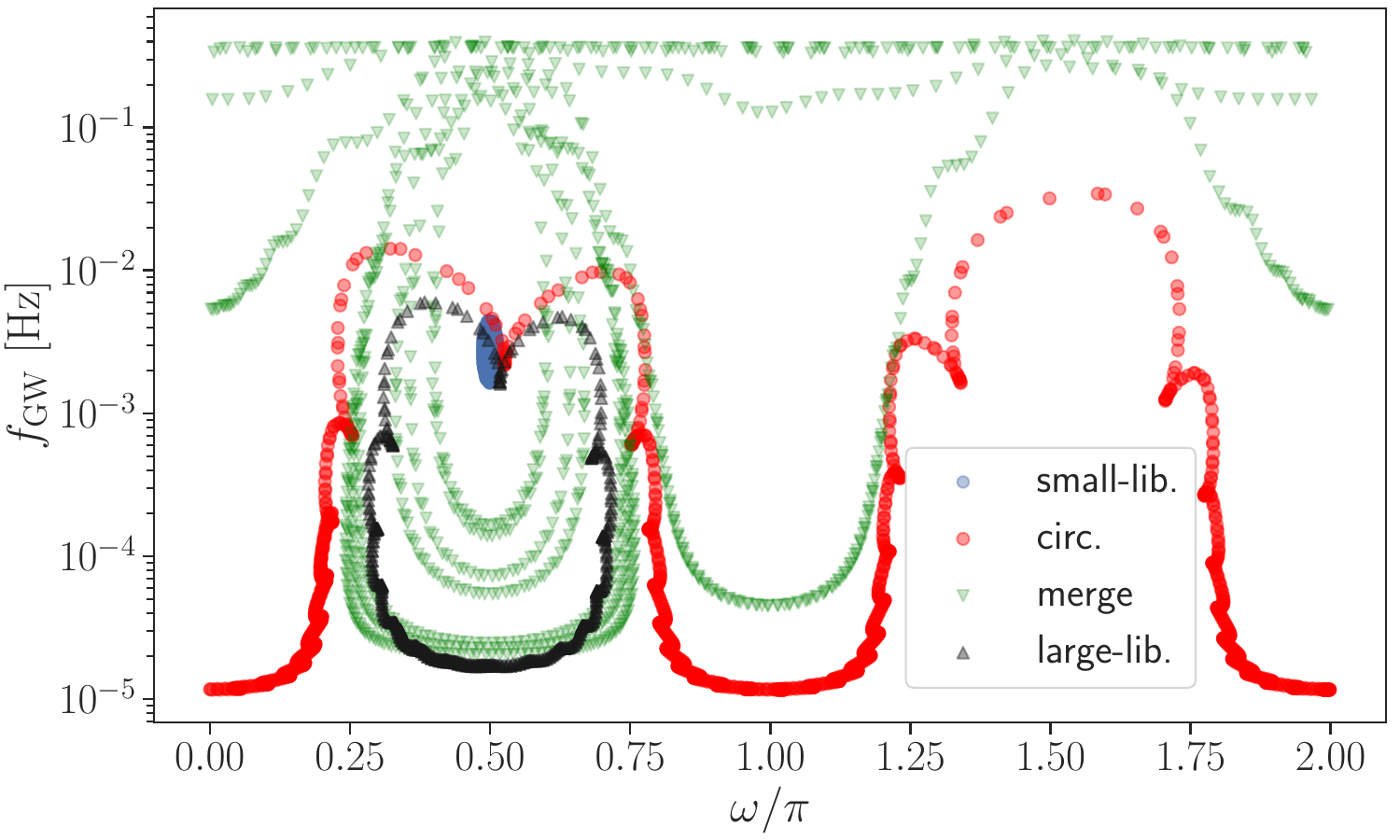}
    \caption{$f_{\rm GW}$-$\omega$ phase portrait for the different orbital families. }\label{all_omega}
\end{figure*}

Mild hierarchies where the inner binary spans a significant fraction of the Hill sphere have extra corrections proportional to the period ratio $\epsilon_{\rm SA}$ in the form of the Brown Hamiltonian \citep{tremaine23}

\begin{equation}
    \epsilon_{\rm SA} = \left(\frac{a_1}{a_2(1-e_2^2)}\right)^{3/2} \left( \frac{M_{\rm SMBH}}{m_0+m_1} \right ) ^{1/2} \approx \frac{P_1}{P_2}. \label{eps_sa}
\end{equation}

The main effect is shifting the locations of the bifurcation points and the extent of the maximal eccentricity, but adding the Brown Hamiltonian does not change the morphology of the phase space. The analytical expression for the fixed points are given by \cite{gri2024a}:

\begin{align}
    e_{\rm fix}=\sqrt{ 1 - \sqrt{ \frac{5(1+\frac{9}{8}\epsilon_{\rm SA}j_z)}{3(1-\frac{9}{8}\epsilon_{\rm SA}j_z)} } |j_z|}. \label{efix}
\end{align}
The associated inclination at the fixed point is
\begin{equation}
    \cos \iota_{\rm mut,fix} = \frac{j_z}{x^{1/2}}= {\rm sign}(j_z) \left( \frac{3}{5} \right)^{1/4} \sqrt{\frac{1-\frac{9}{8}\epsilon_{\rm SA} j_z}{{1+\frac{9}{8}\epsilon_{\rm SA} j_z}} }|j_z|^{1/2}, \label{ifix}
\end{equation}
where $j_z=\sqrt{1-e_1^2}\cos \iota_{\rm mut}$ is a constant of motion and $\epsilon_{\rm SA} \sim P_1/P_2$. At the fixed point, we always have $\omega=\pi/2, 3\pi/2$.

For the highly inclined orbits we consider here, the attained eccentricities are high, and the secular approximation breaks down \citep{antonini12,gri17,gri18}. \cite{f19} shown that direct N-body integrations produce $5-10$ times more mergers (and many eccentric ones) over the secular approach of \cite{hoang2018} for the same initial conditions. \cite{man22} have studied the modified expression for the maximal eccentricities and compared the results with \texttt{TSUNAMI} and found that for the types of orbits considered here where $\epsilon_{\rm SA} \approx 0.05$, the secular approach would fail to capture the highly eccentric bursts. Thus, the direct approach of computing the strain from the accelerations greatly improves upon existing tools of inferring the strain from the secular theory alone \citep[e.g.,][]{knee2024}.
 
\subsection{Numerical integrations and methods}

{\bf In all the following integrations, we use the post-Newtonian $N$-body code \texttt{TSUNAMI} \citep{tsunami-code} to evolve a three-body system. \texttt{TSUNAMI} combines Bulirsch--Stoer extrapolation \citep{stoer1966}, a regularized leapfrog algorithm \citep{mikkola99}, and chain coordinates \citep{mikkola93} to achieve high accuracy and computational efficiency. The extrapolation algorithms uses an adaptive timestep based on convergence rejection with a relative tolerance of $10^{-13}$ \citep{numericalrec}. The code implements post-Newtonian acceleration corrections up to order 3.5 in $v/c$, formulated in the modified harmonic coordinate system \citep{blanchet2024}. Gravitational waveforms are computed from the positions, velocities, and accelerations using the quadrupole formula \citep[e.g.][]{maggiore}, where the accelerations consistently account for the total forces acting on the binary components.}

When estimating the observed GW frequency, we use the \cite{wen03} formula for peak frequency,
\begin{equation}
    f_{\rm GW} = \sqrt{\frac{G(m_1+m_2)}{a_1^3}}\frac{(1+e_1)^{1.1954}}{\pi (1-e_1^2)^{3/2}}. \label{fgw}
\end{equation}

We also calculate the GW strain directly from the total acceleration on each of the binary components using the quadrupole formula (e.g. \citealp{maggiore08}),
\begin{equation}
    h^{\rm TT}_{ab}(t) = \frac{2G}{Dc^4}\ddot{Q}^{\rm TT}_{ab}(t_{\rm ret}).
\end{equation}
Here, TT denotes the transverse-traceless gauge, 
$Q_{\rm ab} = M^{ab} - \delta_{ab}M^{kk}/3$ is the quadrupole moment (the $k$ index is summed over), $t_{\rm ret} = t - D/c$ is the retarded time, and $D$ is the distance from the GW source to the detector. The `TT' superscript stands for the projection onto the line-of-sight via the Lambda Tensor (\citealp{maggiore08}, Eq. 1.36 and 3.59). The mass quadrupole moment is given by 
\begin{equation}
    M^{ab} = m_1 x_1^a x_1^b + m_2 x_2^a x_2^b. \label{mab}
\end{equation}

For the different polarisations and for a given propagation vector described by the angles $\theta,\phi$, the polarisations are generically described by

\begin{align}
h_{+}(t;\theta,\phi) & =\frac{G}{Dc^{4}} \sum_{k=1}^{3}\sum_{l\ge k}^{3}\ddot{M}^{kk}(t)A_{+}^{kl}(\theta,\phi) \nonumber \\
h_{\times}(t;\theta,\phi) & =\frac{G}{Dc^{4}} \sum_{k=1}^{3}\sum_{l\ge k}^{3}\ddot{M}^{kk}(t)A_{\times}^{kl}(\theta,\phi), \label{strains}
\end{align}
where the coefficients $A^{kl}_{+,\times}(\theta,\phi)$ are given by Eq. 3.72 in \cite{maggiore08}. For the second derivative of the mass quadrupole, we have from Eq. \ref{mab}:
\begin{equation}
    \ddot{M}^{ab} = \sum_{j=1}^2 m_j \left(\ddot{x}_j^a x_j^b + 2\dot{x}^a_j\dot{x}^b_j + x_j^a \ddot{x}_j^b\right)
\end{equation}
where the time-derivatives are the instantaneous velocities and accelerations between particles as calculated as in \texttt{TSUNAMI} \footnote{In principle, this term may also include additional accelerations arising from interactions with gas, such as in AGN discs \citep{samsing_nat, pAGN} or during inspiral within a stellar envelope \citep{ginat2020, trani22}, where torque prescriptions are calibrated against hydrodynamical simulations. However, such effects are not relevant for the systems considered in this work.}.

\begin{center}
\begin{table}
\begin{centering}
\begin{tabular}{|c|c|c|c|c|}
\hline 
 & circ. & large lib. & small lib. & merge\tabularnewline
\hline 
\hline 
$e_{1}$ & $0.5$ & $0.5$ & $0.987$ & $0.99877$\tabularnewline
\hline 
$a_{2}$ {[}au{]} & $60$ & $60$ & $60$ & $70$\tabularnewline
\hline 
$\iota_{\rm mut}$ {[}deg{]} & $82.85$ & $82.85$ & $82.85$ & $87.7443$\tabularnewline
\hline 
$\omega\ \rm [deg]$ & $30$ & $90$ & $90$ & $100$\tabularnewline
\hline 
$j_{z}$ & $0.1$ & $0.1$ & $0.02$ & $0.002$\tabularnewline
\hline 
$t_{{\rm end}}$\ {[}yr{]} & $5$ & $2$ & $2$ & $50$\tabularnewline
\hline 
\end{tabular}
\par\end{centering}
\caption{Initial conditions and final integration times for the four orbital families. The BH masses are always equal to $m_1=m_2=20M_\odot$. The inner semi-major axis is always $a_1=0.15\ \rm au$.}\label{Tab1}

\end{table}
\par\end{center}

\section{Orbital families} \label{sec3}

Before going to the detailed dynamics of each orbital family, we provide a brief overview and the key characteristic of each family. The different families represent the possible ZLK dynamics. The initial conditions for each family and the parameter space are studies in detail in \cite{gri2024a}.

In Figure \ref{all_omega}, we plot the correlation between $\omega$ and the GW frequency $f_{\rm GW}$ (Eq. \ref{fgw}) for the different families. Small libration occurs when the orbit is initialised with initial conditions at the fixed point (the small libration due to higher order effects, blue dots). For initial conditions further from the fixed point, but still ones that allow libration in $\omega$, the libration amplitude of $\omega$ and consequently of $e_1$ and $f_{\rm GW}$ is larger (black triangles). For an orbit outside the separatrix, $\omega$ completes a full circulation. The GW frequency becomes LISA-observable around $0.01$~Hz for some limited values of $\omega$ (red dots). For an initially librating orbit with sufficiently large eccentricity, GW dissipation at the pericentre significantly shifts the curves to a new energy curve. Eventually the orbit is tight enough and begins to decouple from the tertiary. At some point $\omega$ starts to circulate and eventually is dominated by GR 1PN precession without changing the eccentricity (green triangles). This orbit regularly achieves GW frequency above $0.1\ \rm Hz$.

In the following we explore each orbital family in more detail and later discuss their observability, rates and future prospects for distinguishing between the different orbits. The initial conditions for all the orbits are $a_1 = 0.15\ \rm au$, $e_2=0.01$, $\omega=\pi/2$,  $\omega_2=\Omega=\Omega_2=0$, $\Omega_2 = \pi$, $\nu_1 = \pi$, $\nu_2 = \pi / 4$. The masses are $m_1=m_2=20 M_\odot$ and $m_3= 4 \times 10^6 M_\odot$. Table \ref{Tab1} shows the other parameters that vary between different simulations. We also specify the initial value of $j_z=\sqrt{1-e_1^2}\cos \iota_{\rm mut}$, and the final time of the simulation. Each initial condition is a representative of a typical family of orbits.



\begin{figure*}
    \centering
    \includegraphics[width=0.93\textwidth]{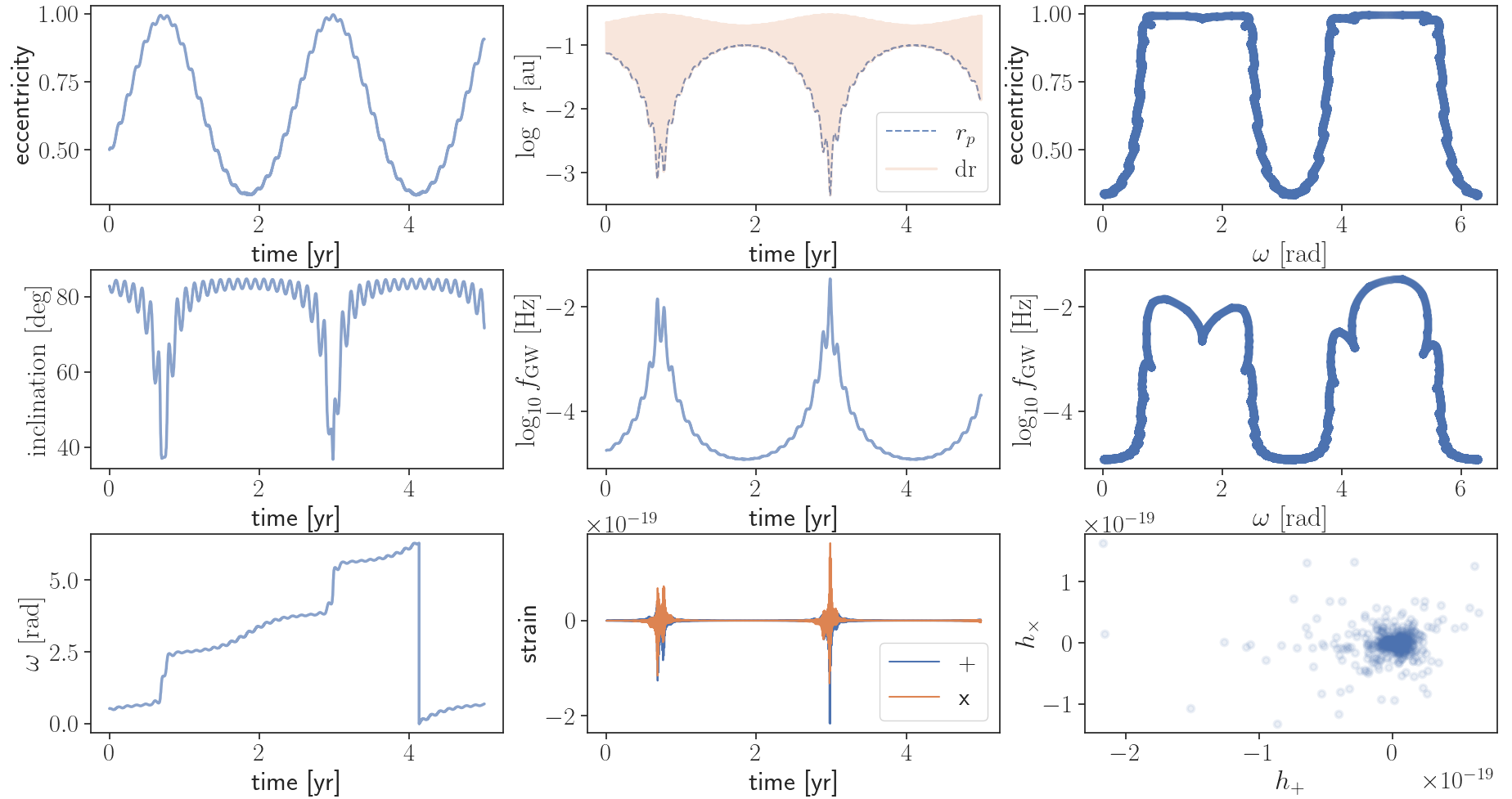}
    \caption{Representative of the circulating family of orbits. The initial conditions are $m_1=m_2=20$~M$_\odot$, $a_1 = 0.15\ \rm au$, $a_2=60\ \rm au$, $e_1 = 0.1$, $e_2=0.01$, $\iota_{\rm mut} = 82.85\ \rm deg$, $\Omega_2=\omega_2=\Omega=0$, $\Omega_2=\nu_1=\pi$ and $\nu_2=45\ \rm deg$. inner orbit at fixed point, $a_1=0.15$ au, $j_z=0.1$. Left column (top to bottom) eccentricity, mutual inclination and argument of pericentre versus time. Middle column: instantaneous separation $\rm dr$ and pericentre $r_p$, GW frequency $f_{\rm GW}$ and the strain polarisations versus time. Right column: scatter plots of eccentricity versus $\omega$, GW frequency versus $\omega$ and the two strain polarisations. \label{circ}}
\end{figure*}

\begin{figure*}
    \centering
    \includegraphics[width=0.93\textwidth]{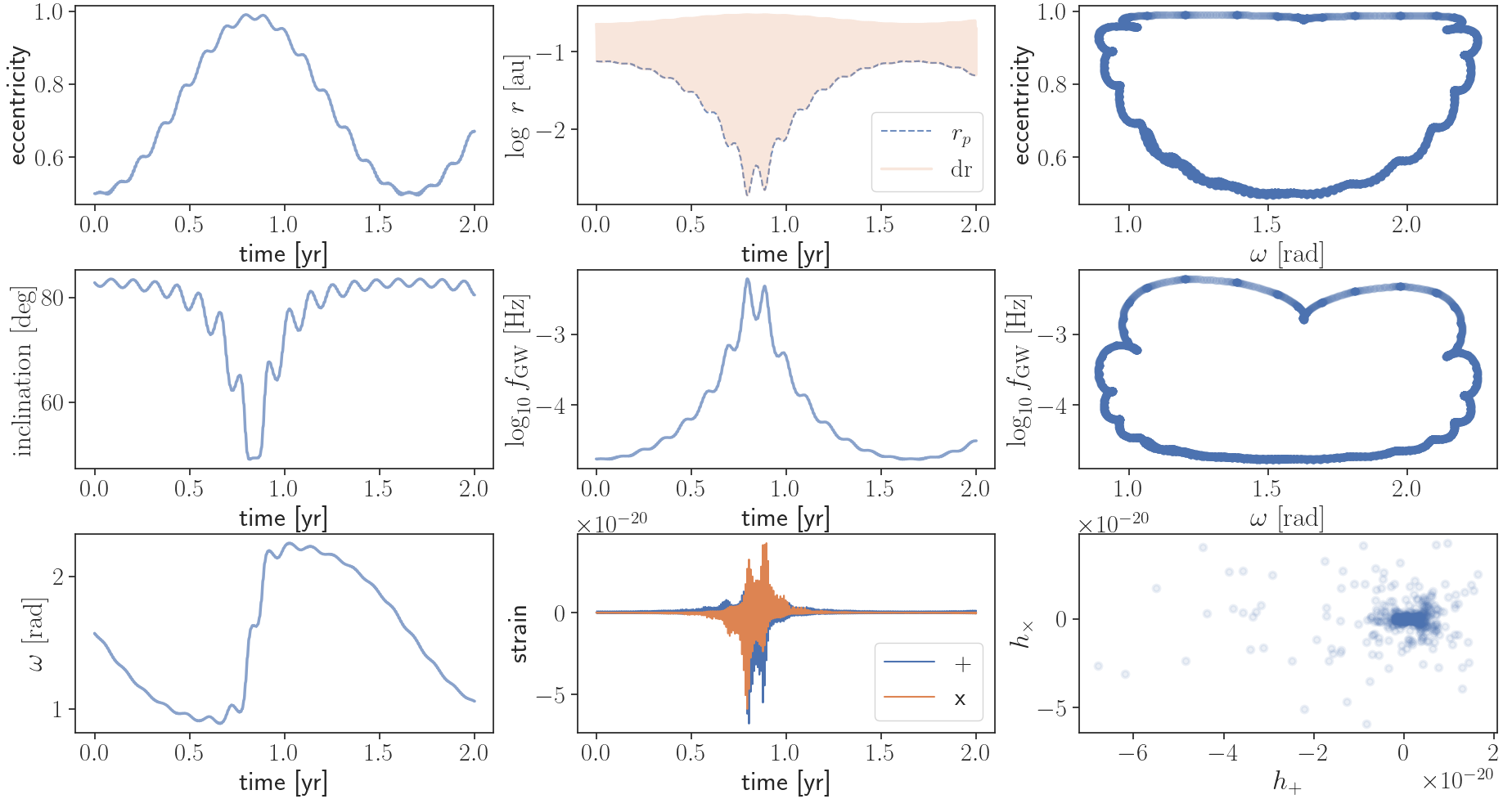}
    \caption{Representative of the large amplitude libration orbital family. Similar to Figure \ref{circ}, but with different initial conditions (see table \ref{Tab1}).}\label{large_lib}
\end{figure*}

\begin{figure*}
    \centering
    \includegraphics[width=0.99\textwidth]{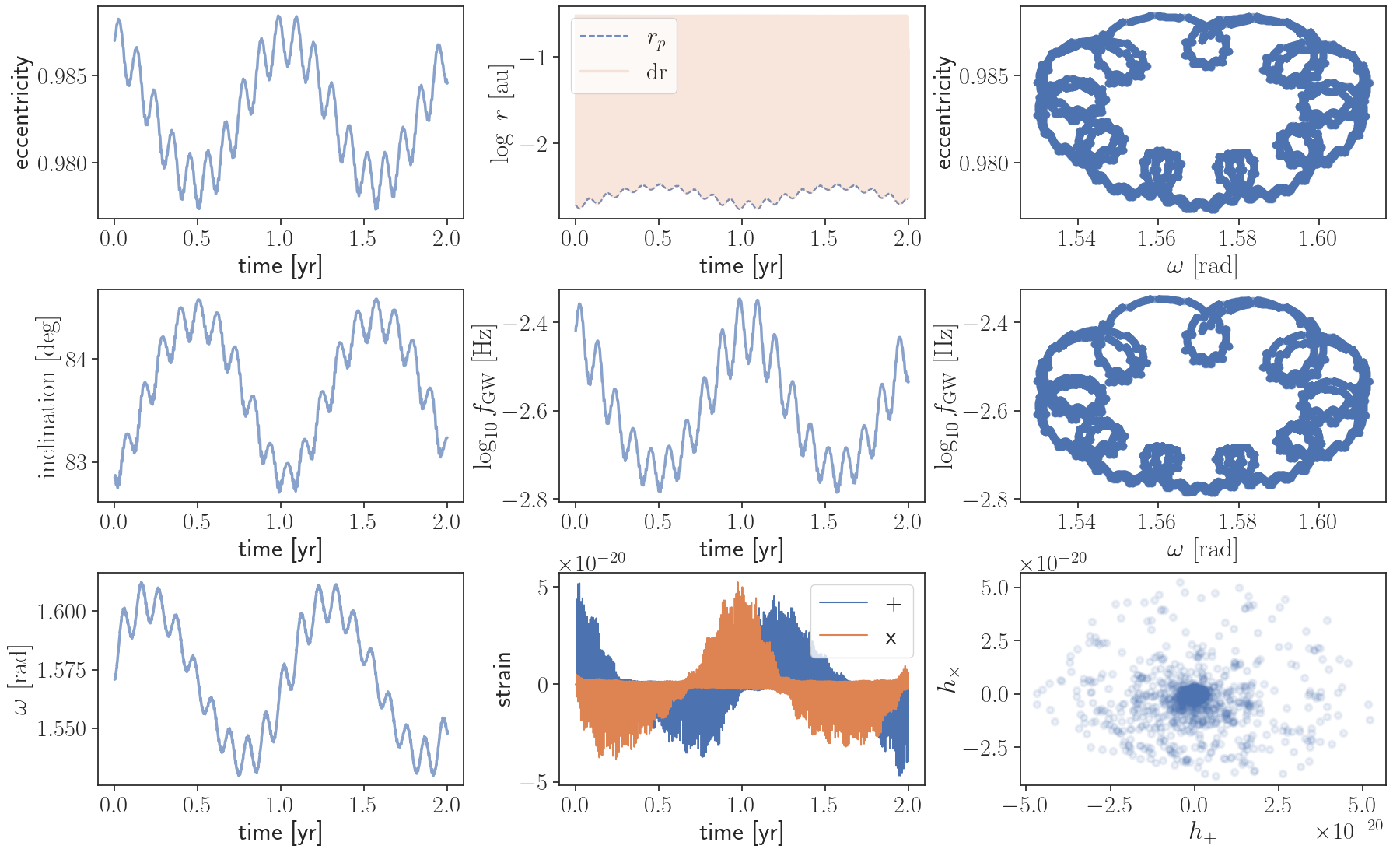}
    \caption{Representative of the small amplitude libration family, similar to Figs. \ref{circ} and \ref{large_lib}. Initial conditions are near the fixed point in phase space.}\label{smal_lib}
\end{figure*}

\begin{figure*}
    \centering
    \includegraphics[width=0.95\textwidth]{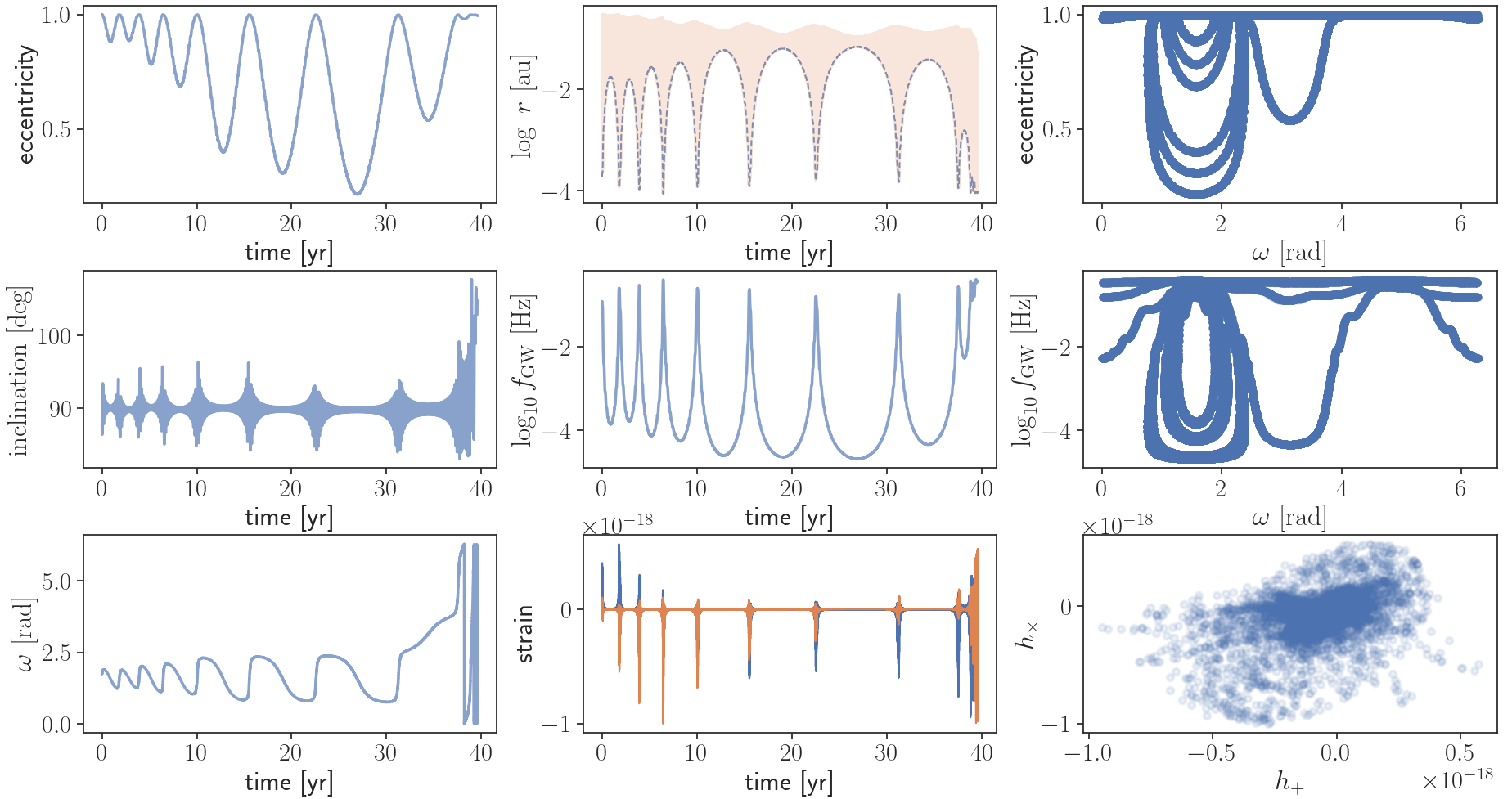}
    \caption{Representative of the librating and merging family of orbits. The panels are similar to Figure \ref{circ}. The initial conditions are described in table \ref{Tab1}.}\label{merge}
\end{figure*}

 \begin{figure*}
    \centering
    \includegraphics[width=0.94\textwidth]{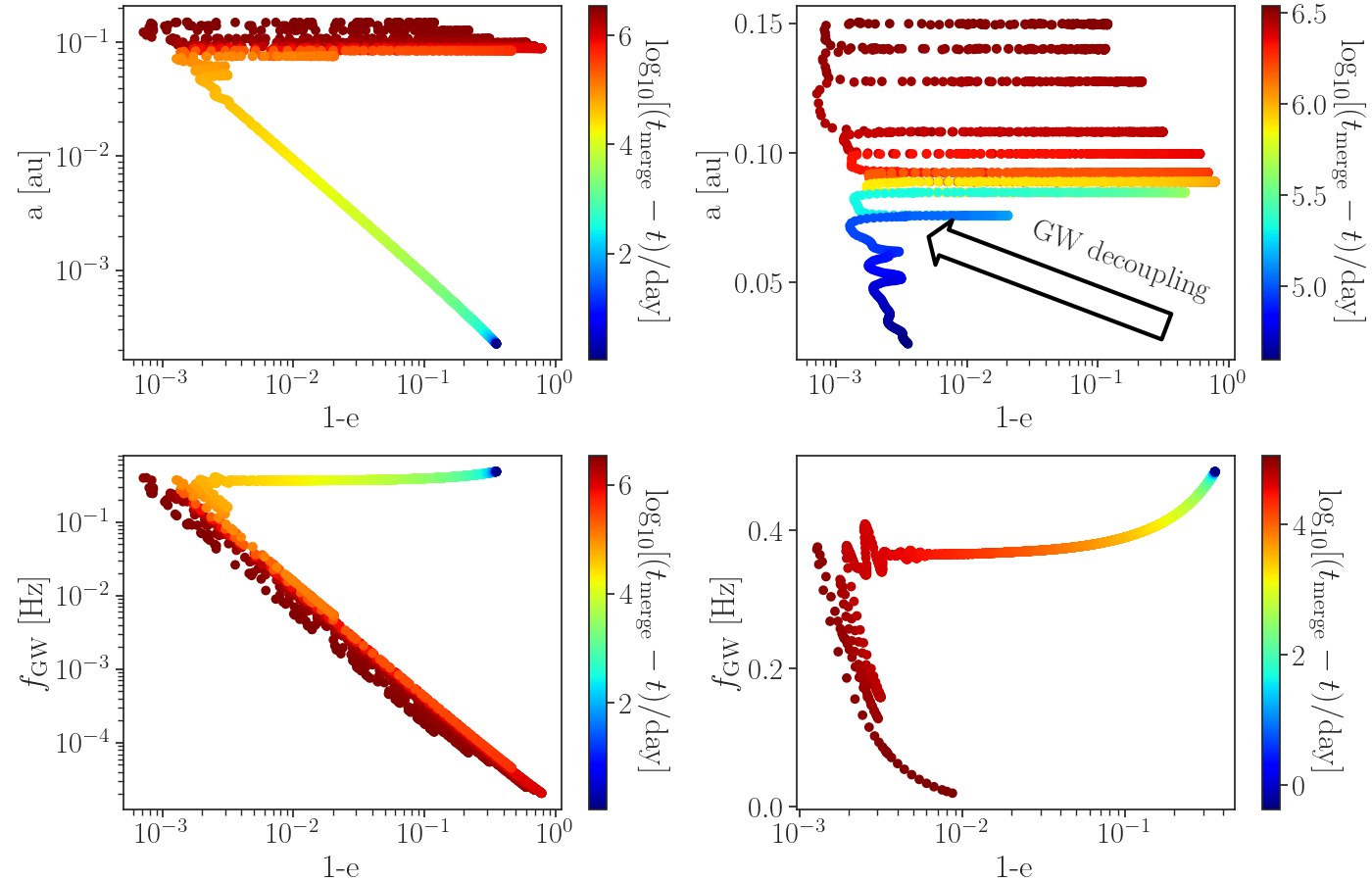}
    
    \caption{Evolution of the librating and colliding orbit. Each cycle slightly decrease the separation until the binary is captured and decoupled from the tertiary. The right panel shows a zoom-in version on the last ZLK cycle, capture and inspiral.}\label{merge_detailed}
\end{figure*}

\subsection{Circulating orbits}

The only example of a population of highly inclined hierarchical triples with extreme mass ratios is the population of irregular satellites around giant planets in the Solar System. Out of the $228$ irregular satellites\footnote{By irregular we mean satellites with orbit beyond the Laplace radius \citep{tremaine23}, which are believed to be formed by capture.}, only $13$ are found in ZLK libration, the rest are circulating \citep{gri2024b}. Theoretically, the area in $e_1$-$\omega$ phase space is larger for circulating orbit, with the actual ratio depending on $j_z$. Thus, the circulating family of orbits is the most common.  Figure~\ref{circ} depicts the orbital evolution of a representative orbit. The orbit starts from a relatively low/moderate eccentricity (in this case $e_1=0.5$, top left) and grows to large values. The mutual inclination is sharply decreasing in a correlated manner that preserves $j_z$ (on average; middle left). The argument of pericentre $\omega$ is circulating and completes a full revolution (bottom left).

The top middle panel shows the instantaneous separation ${\rm d}r$ and the pericentre $a_1(1-e_1)$ which also changes by almost three orders of magnitude. The central panels show the evolution of the peak frequency $f_{\rm GW}$ with time. The peak frequency grows by almost three orders of magnitude for the circulating family. The middle bottom panel shows both strain polarisations $h_+$ and $h_\times$ for $D=8\ \rm kpc$, $\theta,\phi=0$ (head-on GW propagation). We see that the strain is highly bursty and peaks to  strain magnitudes of about $10^{-19}$ for a short duration during the highly eccentric phase.

The right panels show the correlation between different parameters. The phase portrait which is typical for ZLK orbits, where $\omega$ quickly crosses $\pi/2, 3\pi/2$ at high eccentricity. This is because the secular evolution is proportional to $t_{\rm sec} / \sqrt{1-e_1^2}$. The peak of $f_{\rm GW}$ (which occurs at the same time as the peak of $e_{\rm max}$) is associated with the time at which $\omega$ is hovering around integer multiples of $\pi/2$. This is a generic feature of binaries of high mutual inclination in hierarchical triple systems \citep{hamers2021}. The peak frequency is not identical and depends on the exact value of $e_1$ at the peak, which is stochastic in nature when the orbit-averaging solution breaks down \citep[see e.g.,][]{gri18,man22}. Finally, the bottom right panels shows the correlation between the two strain polarisations.

\subsection{Large amplitude librating orbits}
Figure \ref{large_lib} shows an example of an orbit that librates with a large amplitude $\omega$ between $\sim 0.75$ and $2.25$ rad, a range of almost $\pi/2$ radians (or 90 degrees) (top right). The only difference between this orbit and the circulating orbit is a different starting point of $\omega$. The peak frequency $f_{\rm GW}$ also changes by over two orders of magnitude, and the maximum peak is again achieved at $\omega \approx \pi/2$. The phase portrait of $e_1$-$\omega$ is now on a closed loop, but the behaviour is similar to the circulating orbit where $\omega$ sweeps quickly in the $e_1 \sim \omega$ phase space. This is associated with the greatest loss of GW radiation, which is also indicated in the peak of the strain polarisations. Overall the large libration orbital family is similar to the circulating one. This is because these orbits are relatively close to the separatrix, which is the critical curve that separates circulating and librating orbits and crosses the unstable (hyperbolic) fixed point of $e_1=0, \omega=0$, similar to the pendulum. (see details e.g., in \citealp{gri2024a}). 

\subsection{Small amplitude librating orbits}
Figure \ref{smal_lib} shows the orbital evolution of a small amplitude librating orbit. We see that the orbital elements vary little, and the variation in $f_{\rm GW}$ varies by $0.4$ dex. The argument of pericentre is confined to a small range of $\sim 7.5$ degrees. The system stays at relatively high eccentricity, and the waveform has relatively regular bursts on the inner period timescale $P_1 \approx 0.0092\ \rm yr$, which is different from the other systems that experience quiescent phases over the secular timescale of $\sim 1-2\ \rm  yr$. The phase portrait shows clear motion of small librations around a fixed point. We will examine the small amplitude libration orbit, the different timescales involved (inner orbit $P_1 \ll$ outer orbit $P_2 \ll$ secular time $t_{\rm sec}$ and the comparison to an isolated binary evolution in sec. \ref{4.3}.

\subsection{Merging orbits}
Figure \ref{merge} shows the orbital evolution of the merging orbit, which is the most varied of the orbital evolution families and has several phases. Initially, the orbit is in libration. However, due to the initial large eccentricity, each pericentre passage \textit{increases} the libration amplitude: the energy decay changes the locations of the fixed point and detunes the system, making the libration amplitude larger. The decreasing semi-major axis causes the secular period to increase, and eventually, after $\sim 6$ cycles and about $\sim 30$ years of orbital evolution, $\omega$ escapes and starts to circulate. After another cycle, GW capture occurs and the GW dissipation is strong enough to reduce the semi-major axis such that the inner binary is essentially decoupled from the central BH. 

\begin{figure*}
    \centering
    \includegraphics[width=0.99\textwidth]{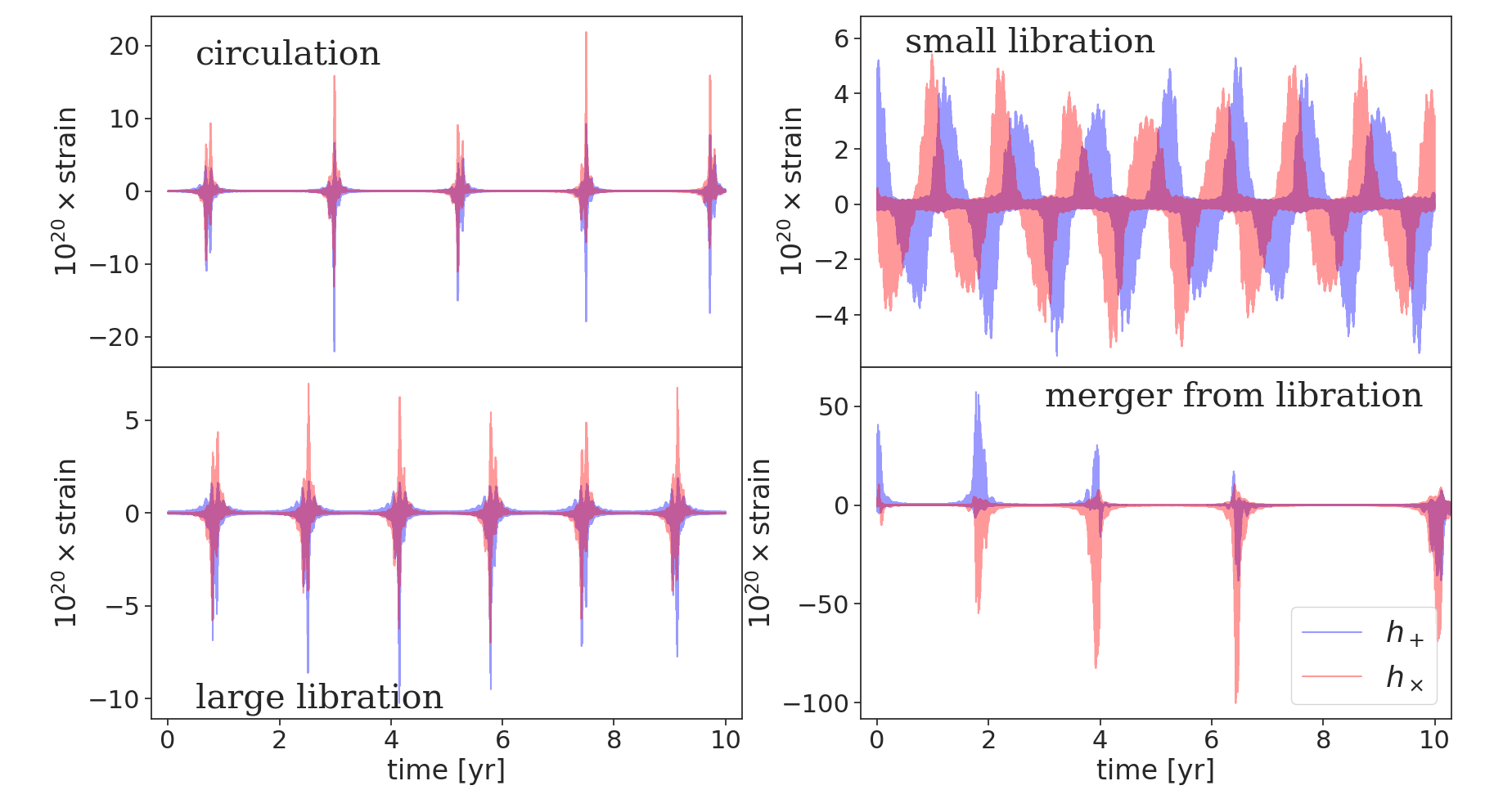}
    \caption{Dimensionless strain polarisations for the four orbital families initialised as in table \ref{Tab1} for a distance of $8\ \rm kpc$. The time between the evaluations is $\Delta t = 10^4\ \rm s$ (however it can be shorter for a close approach). The total integration time is 10 yrs. The left panels show the circulation and large libration orbits. The right panels show the small libration and merger orbits. The purple and pink colour show the plus and cross polarisation, respectively.}\label{fig8}
\end{figure*}

Finally, the inner binary merges. Using \citet{Peters64} to evolve the binary from its final recorded state at $0.26$~Hz, we estimate that the eccentricity of this binary at $10$~Hz 22-mode GW frequency is $0.016$. This may be slightly below the current threshold of LVK detectors \citep[$0.05$ at $10$~Hz for GW150914-like systems detected with third observing run sensitivity;][]{lower18}, however, lighter systems detected with increasingly sensitive detectors may have a lower minimum detectable eccentricity. Future third-generation detectors with improved sensitivity and lower minimum frequencies, expected in the mid-2030s and coinciding with LISA, will improve eccentricity measurement sensitivity significantly \citep{Saini:2024:ETsensitivity}. At $1$~Hz ($5$~Hz) 22-mode GW frequency this binary retains an eccentricity of $0.172$ ($0.033$). The rapid variations of $\omega$ observed in Figure \ref{merge} are due to fast GR precession from the $1$PN term, which is associated with the increase of the GW frequency, and potentially observable in ground-based GW detectors. We also note that the final inclination with respect to the outer orbit is retrograde, exceeding $100$ deg.

Figure \ref{merge_detailed} shows the evolution of the merging orbit in the $a-e_1$ parameter space (top panels) and in $f_{\rm GW}-e_1$ space (bottom panels), colour coded by the time before merger $t_{\rm merge} - t$. We see that initially the binary is at fixed $a$ and emits a burst of GW radiation every time it becomes highly eccentric, near $1-e \sim 10^{-3} \times (0.6-2)$, causing it to experience a step-like decrease in separation. After about $\mathcal{O}(10)$ cycles, the binary eventually decouples from the tertiary SMBH. 

The decoupling occurs where relativistic effects take over the ZLK oscillations: The first instance occurs at $a_1 \approx 0.075$ and $e_1=1-0.0015$, where GW dissipation timescale for an eccentric binary $t_{\rm GW}(a_1,e_1)$ is comparable to the timescale of eccentricity oscillation $t_{\rm sec}\sqrt{1-e_{\rm 1}^2}$, which is faster than the secular ZLK timescale $t_{\rm sec}$ in the highly eccentric phase \citep[e.g.][]{gri18}. In the next oscillation $a_1$ decreases $a_{\rm 1} \approx 0.056\ \rm au$ and eccentricity $1 - e_{\rm 1}\sim 0.0018$, after this point the 1PN term, $\epsilon_{\rm GR} \approx 0.4$ (Eq. \ref{eps_gr}), dominates and the 1PN precession begins to quench ZLK oscillations.

The two effects start to dominate over the ZLK oscillations roughly at the same time. Using the \cite{Peters64} formula for the GW timescale at the 2.5PN level and the timescale for the 1PN precession, the timescale ratio is 

\begin{align}
    \frac{t_{{\rm GW}}}{t_{{\rm 1pn}}}	& \approx\frac{9}{64\pi} \left(\frac{a_1}{r_g} \right)^{3/2} (1-e_1^{2})^{5/2} \\ \nonumber
	& =4\times10^{6}\left(\frac{a_1}{0.08\ {\rm au}}\right)^{3/2}(1-e_1^{2})^{5/2}. \label{tratio}
\end{align}
For our points of interest $a_1/{\rm au} = \{0.075, 0.056 \}$ and $1 - e_1=\{0.0015, 0.0018 \}$ we get that the ratio is $t_{\rm GW}/t_{\rm 1pn} \approx \{0.96, 1.8 \}$, respectively, so both effects are comparable for a highly eccentric orbit. As the binary inspirals further and becomes more circular, the 1PN effects will become faster, so $\omega$ is expected to complete many revolutions before the BBH merges.


Inspiralling stellar-mass BBHs sweep through the LISA band several years before they merge, entering at detectable amplitudes at ${\sim}0.01\ \rm Hz$ \citep[e.g.,][]{Klein:2022:lastthreeyears}. These binaries, if within the LISA observational distance volume, would exhibit bursts of GW emission at its highest-eccentricity phases throughout its life. If observed during its inspiral phase, either in real-time or through an archival search, it would be observed with initially near-maximal eccentricity as a burst-like source. It would be identifiable using methods like the ones outlined in \citet{knee2024}, and the parameters of the system could be identified using burst timing inference methods \citep{rs2023}. Implementing extraction of burst time information from \texttt{TSUNAMI}, and creation of an approximate model from this (since full waveform generation takes ${\sim}\mathcal{O}(1)$~s currently and is thus unsuitable for performing inference), is intended for future work.

\section{Detectability and astrophysical implications} \label{sec4}

\subsection{Distinguishing between the orbits}

By slightly varying the initial conditions, we identify four families of orbits. The rich parameter space, the chaotic nature of the problem, and the time that waveform generation takes mean that, at this stage, full parameter estimation of these sources is not feasible. However, given the future LISA mission, we are able to draw some preliminary conclusions. 

\begin{figure*}
    \centering
\includegraphics[width=0.96\textwidth]{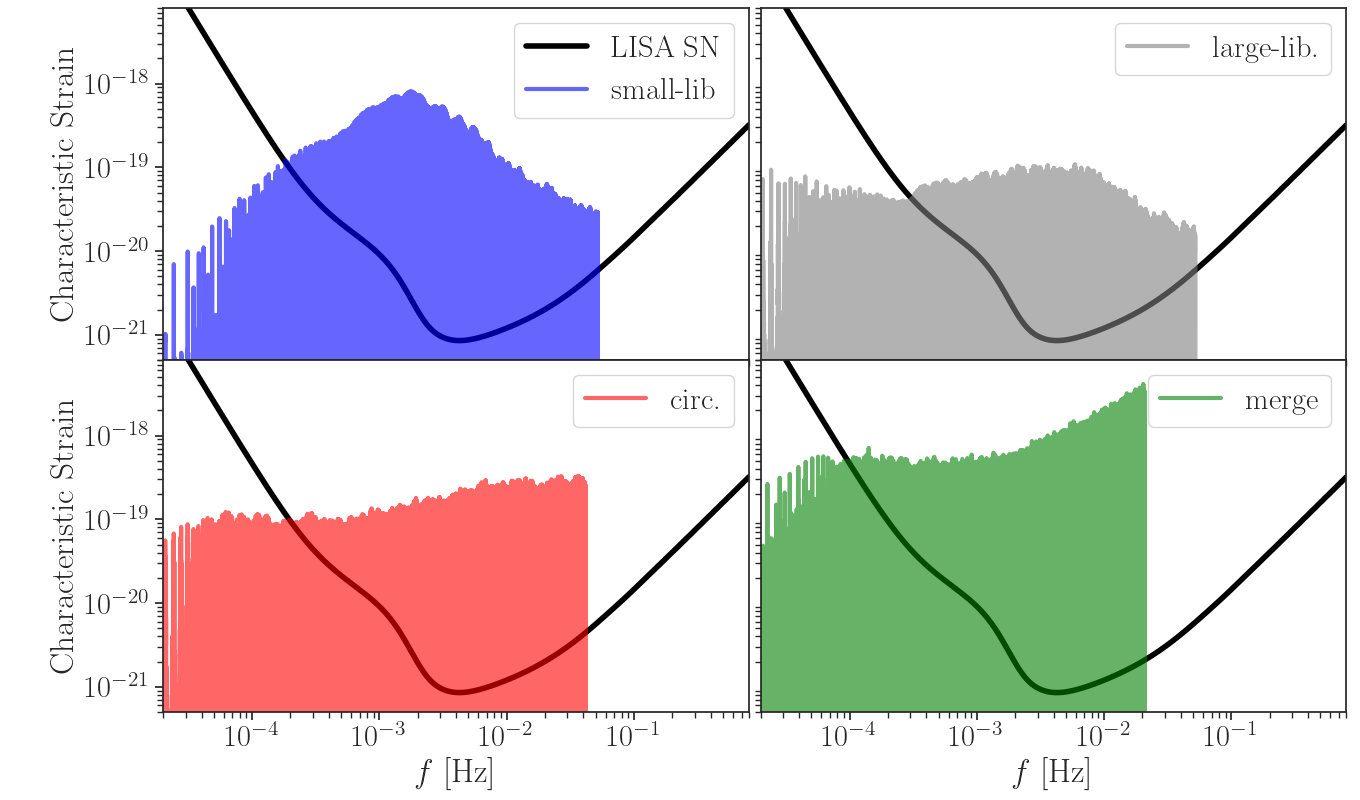}
    \caption{Characteristic strain $h_c=2f|\tilde{h}(f)|$ of the four orbital families overlaid with the LISA's sensitivity curve (thick black line). The signals are here curtailed artificially suddenly at the highest-frequency point in the simulation output. }\label{char}
\end{figure*}

The merging orbit is the most probable observable GW source. This orbit starts on a low value of $j_z$, which leads to a high eccentricity and corresponding $f_{\rm GW}$. In the first few cycles of the GW capture, $\omega$ rapidly circulates due to the 1PN term; this occurs at $f_{\rm GW}\approx 0.3\ \rm Hz$. The simulation stops at $f_{\rm GW} \approx 0.48\ \rm Hz$ at $a_0$ and $e_0$, as in Figure \ref{merge_detailed}. The merging orbit may be observed in LISA first as a high-eccentricity burst source, before being captured inspiraling through LISA and into ground-based GW detectors. This source may be observed in real-time in LISA, or alternatively may be observed with ground-based detectors before being searched for in archival LISA data. 

We compare the different strains obtained for our four representative orbits in Figure \ref{fig8}. We use the same sampling rate of $10^{-4}$~Hz and the same time of integration of $10$ years. We see that for the orbits that exhibit large eccentricity variation---i.e, all orbital families except small libration---the strain has distinct sharp bursts and reaches strain amplitude magnitudes up to $\approx10^{-19}$, and higher for the merging binary (for the Galactic distance of $8 \ \rm kpc$). The time difference between the peaks is the secular time, which is slightly different between the orbits. The merging orbit has a varying secular timescale due to the changing libration amplitude (the capture phase is not shown). The small librating orbit does not exhibit strong bursts and is more sinusoidal throughout the evolution.

Figure \ref{char} shows the characteristic strain of the four families overlaid with the LISA sensitivity curve. The characteristic strain is $h_c(f)=2f|\tilde{h}(f)$|, where $\tilde{h}(f)=\sqrt{\tilde{h}^2_x(f) + \tilde{h}^2_+(f)}$ is the sum of the Fourier components of the two strain polarisations. For the Fourier transform we use the fast Fourier Transform implemented within the \texttt{BILBY} package \citep{ashton19}, where \texttt{scipy.interpolate}'s interpolator was used to construct a uniform time series from the simulated strains. LISA's sensitivity curve (thick black line) was generated using \cite{lisaSn}'s Jupyter notebook. 

We see that all orbital families have a distinctive detectable features where the peak strain exceeds the LISA sensitivity curve by about two orders of magnitude near the minimum of the sensitivity curve around $f_{\rm GW} \approx 3-5 \times 10^{-3}\ \rm Hz$. The merging orbit remains high throughout the LISA frequency band. These preliminary high SNRs make future detections potentially feasible.

These waveforms are generated numerically using the new functionality we have added to the \texttt{TSUNAMI} code. If the time between output snapshots is $\Delta t$, then each waveform takes $\mathcal{O}(1) \times (1+\Delta t / 5000\ \rm s)$~s to generate, which makes Bayesian inference studies infeasible. In future work, we will extract burst time information from \texttt{TSUNAMI} and create an approximate surrogate model to enable the parameters of the systems to be inferred \citep[e.g.,][]{rs2023}. 

\subsection{Total number of librating BBH}
The rates of the systems we investigate in this paper are uncertain. Although about $10^4$ BHs are expected to reside within a distance of $10^3\ \rm au$ from Sgr A* \citep{me00,freitag06,rose22}, it is unclear how many of them are in binaries or in librating configuration. The only large statistical sample of well resolved orbits of similar configurations of small-mass binaries orbits a much more massive object is the sample of irregular satellites in the Solar system, where $228$ irregular satellites around giant planets are known, and only $13/228\approx 5.7\%$ are found to be librating \citep{gri2024b}. However, these orbits are markedly different with relative low inclinations with $|j_z| >0.4$, so they do not experience high eccentricities. There is also a distinct void in the range of mutual inclinations $60-120^\circ$, attributed to the loss of the satellite during ZLK cycles \citep{gri17}.

The conditions for libration to be possible is given by the dimensionless quantities $\epsilon_{\rm SA}$ (Eq. \ref{eps_sa}) and $j_z$ (see Figure 7 of \citealp{gri2024a} for 'optimal' libration zone where the initial $\omega=\pi/2$ and Figure 1 of \citealp{gri2024b} for the sample of known irregular satellites), however they are inaccurate since they neglect the dependence of $\omega$.

Recently, a high-order analytical theory was developed to classify librating orbits with the extent of the Brown Hamiltonian (accounting for mild hierarchies, \citealp{lg25a, lg25b}). In particular, the orbit will be librating if the modified 'Kozai constant' 
\begin{equation}
    C_{\rm ZLK} = e^2 \left[1 + \frac{9}{2}\epsilon_{\rm SA} j_z - \frac{5}{2}\left(1 + \frac{9}{8}\epsilon_{\rm SA}j_z \right) \sin^2\omega \sin^2\iota \right]
\end{equation}
is negative (Eq. 20 in \citealp{lg25b}, see also their Figure 2 for parameter space exploration). For $\epsilon_{\rm SA}=0$, $C_{\rm ZLK}$ reduces to the classical ZLK constant.

\begin{figure}
    \centering
    \includegraphics[width=0.49\textwidth]{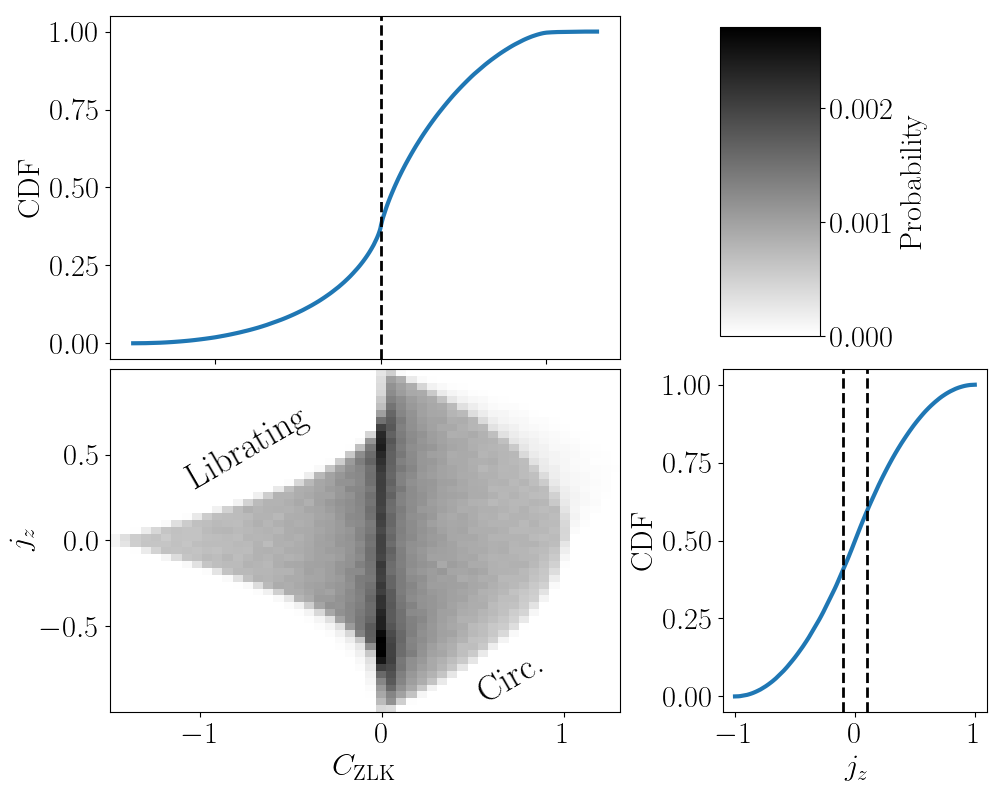}
    \caption{Parameter space of $C_{\rm ZLK}$-$j_z$ that distinguishes circulating ($C_{\rm ZLK}>0$) and librating $(C_{\rm ZLK}<0)$ orbits from our fiducial sample. About $37\%$ of the orbits are librating (top panel CDF). Another $\sim 20\%$ has low $j_z$ (bottom panel CDF).}\label{pop_zlk}
\end{figure}

\begin{figure*}
    \centering
    \includegraphics[width=0.99\textwidth]{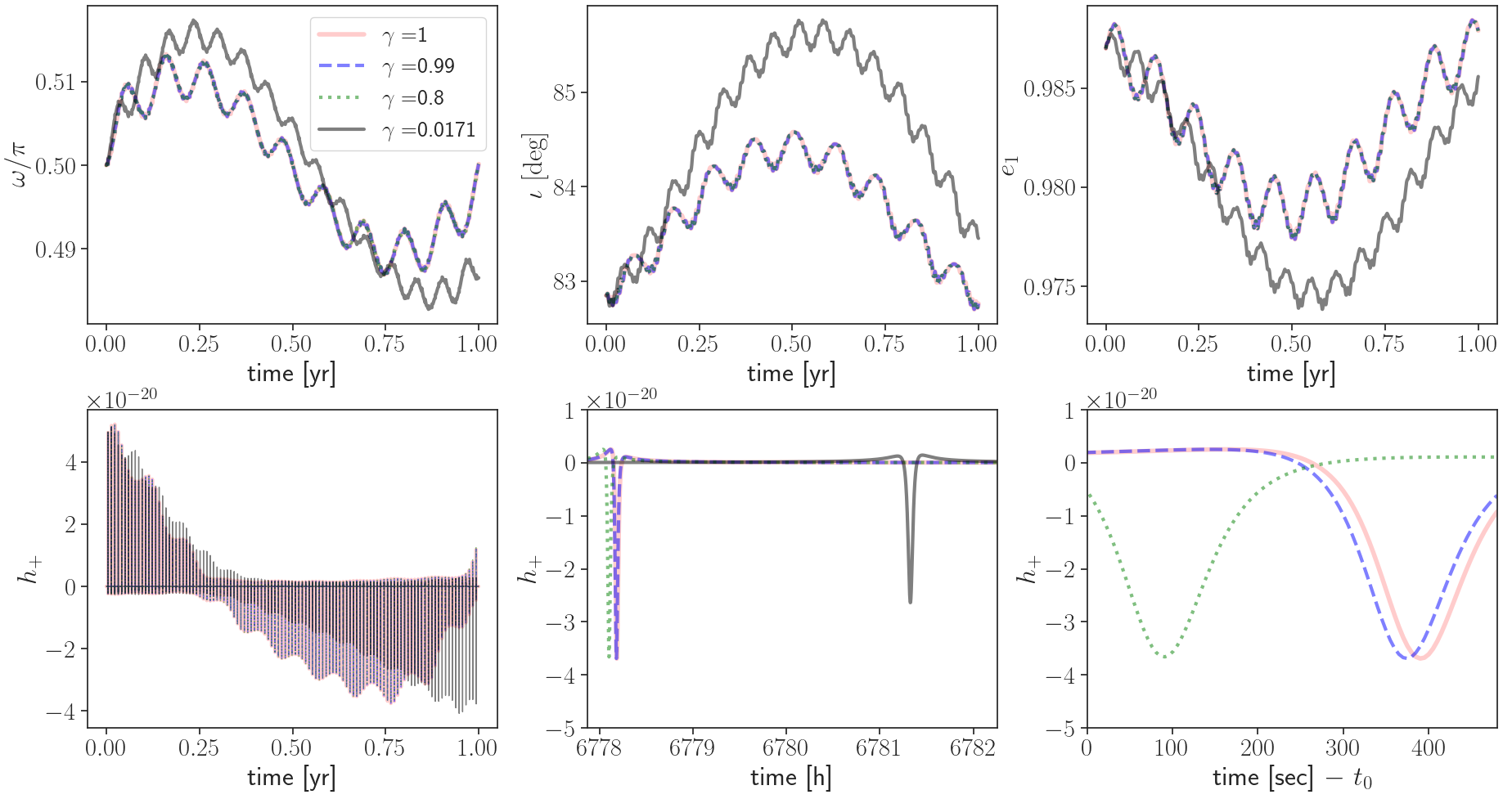}
    \caption{Time evolution of the small libration orbit for different values of $\gamma$, where the tertiary mass and separation are scaled as $m_{\rm out} \to \gamma^3 m_{\rm out}$ and $a_2 \to \gamma a_2$, which preserves the secular time. The pink solid line is the standard orbit $\gamma=1$, dashed blue is the closest orbit $\gamma=0.99$. Dotted green is $\gamma=0.8$,  solid black is the orbit with three equal masses, where $m_{\rm out}=m_1=m_2=20\ M_\odot$ ($\gamma=0.0171)$. The orbital elements are shown in the top panels, while the $'+'$ strain polarisation is shown in the bottom panels. Bottom middle and right panels are zoomed in version on timescales of hours and minutes, respectively.}\label{fig9}
\end{figure*}

BBHs, of course, have very different formation and evolution channels. If the BBH binary fraction is $f_{\rm bin}$, and the mutual inclination is uniform in cosine, we may expect $\sim 10^4 f_{\rm bin} f_{\rm high,lib}$ where $f_{\rm high,lib}$  is the number of librating binaries with $|j_z|\le 0.1$ such that hte eccentricity is large.

In order to explore these fractions, we take $N=10^6$ random samples of a triple system with $\omega \sim U[0,2\pi]$ sampled from a uniform distribution, $e$ sampled from a thermal distribution $(f(e)=2e). \cos \iota \sim U[-1,1]$ also uniform and $\log_{\rm 10} \epsilon_{\rm SA} \sim U[-3,-0.5]$ is sampled from a log-uniform distribution corresponding to \"{O}pik's law.

Figure \ref{pop_zlk} shows the distribution of the orbits in $j_z$-$C_{\rm ZLK}$ space, which is very similar to the parameter space presented in Figure 2 of \cite{lg25b}. We see that $37\%$ of the orbits are librating $(C_{\rm ZLK}<0)$, $\sim 20\%$ have $|j_z|>0.1$. The total number of highly-eccentric librating orbits (both $C_{\rm ZLK}<0$ and $|j_z|<0.1$) is about $10\%$. For the irregular satellite population, about $7\%$ of the orbits are librating with $|j_z|>0.4$, compatible with the observations. Fixing $\epsilon_{\rm SA}=10^{-0.5}$ only raises the fraction of librating orbit to $\sim 42\%$ but the fraction of highly-eccentric librating orbits remains similar, thus we conclude that our studies sample does not strongly depend on the prior of $\epsilon_{\rm SA}$.

To summarise, $f_{\rm high,lib}\approx 0.1$ and we expect about $10^3f_{\rm bin}$ highly eccentric librating binaries. The same estimate for low eccentricity librating binaries is about $7\%$, compatible with the  $5.7\%$ of the librating irregular satellite sample. The binary fraction is unknown, but BHs do tend to pair up efficiently in nuclear star clusters \citep[e.g.,][]{2023PhRvD.108h3012K, 2023MNRAS.526.4908C, 2024A&A...683A.186D}.

\subsection{Generalisation to arbitrary tertiary mass}

The inner binary is perturbed by a massive distant tertiary of mass $m_{\rm out}$ at separation $a_{\rm out}$. The typical timescale of this perturbation is the secular timescale \citep[e.g.,][]{gri2024a},

\begin{equation}
    t_{\rm sec} = \frac{4(m_1+m_2)^{1/2}a_1^3(1-e_2^3)^{3/2}}{3G^{1/2}m_{\rm out}a_1^{3/2}} \propto \frac{a_2^3}{m_{\rm out}}.
\end{equation}
For a given scalar $\gamma > 0$, scaling the mass by $m_3 \to \gamma^3 m_3$,  and the separation by $a_2 \to \gamma a_2$ keeps the secular time, and thus the strength of the perturbation intact.

The secular approximation yields the orbital evolution on secular times, after averaging over short term oscillations. Recent studies have used the secular approximation to study the detectability of eccentric BBH in galactic nuclei \citep{hoang19, randall19, knee2024}. Another approximation is the test particle limit, where the ratio of the inner to outer angular momentum, is small, i.e. 
\begin{align}
\eta \equiv  \frac{\mu_{\rm in}}{\mu_{\rm out}}\bigg[ \frac{m_{\rm bin}a_1}{m_{\rm tot}a_{2}(1 - e_{\rm out}^2)}\bigg]^{1/2},
\label{eq:eta}
\end{align}
where $\mu_{\rm in} = m_1 m_2/m_{\rm bin}$ and $\mu_{\rm out} = m_{\rm bin} m_3 / (m_3 + m_{\rm bin})$ are the reduced masses of the inner and outer binaries, respectively. For the small libration case, $\eta_0=3.95 \times 10^{-5}$. The angular momentum ratio scales as $\eta \to \gamma^{-2} \eta$.  Even within the framework of the double-averaging (secular) approximation, values of the angular-momentum ratio as low as $\eta \approx 10^{-4}$ could yield significant changes in the secular evolution \citep{naoz13}.

Figure \ref{fig9} shows the evolution of the small libration orbit for different values of $\gamma$. We see that the equal mass case ($\gamma=0.0171;\ m_{\rm out}=\gamma^3 \times 4\times10^6 \approx 20\ M_\odot$) is evolving significantly differently, as seen in the orbital elements in the top panels. The difference between the burst timing  (bottom middle panel) is about $3$ hours. The other orbits ($\gamma=0.8, 0.99$) evolve similarly to the fiducial orbit, and the orbital element evolution is indistinguishable after 1 year. However, since the width of each burst is narrow, the GW strain signature is still different. The bottom right panel shows that the offset between the $\gamma=0.8$ orbit and the fiducial orbit is about $4-5\ \rm min$, while the typical width of the burst is about $2\ \rm mins$, so the bursts are largely separate from each other. Only the $\gamma=0.99$ orbit significantly overlaps with the fiducial orbit, with an offset of about $20\ \rm s$. However, one year of data contains over $100$ peaks (since the inner orbital period is slightly less than $0.01\ \rm yr$). Even orbits with small differences in $\gamma$ are therefore expected to be distinguishable.

\begin{figure*}
    \centering
    \includegraphics[width=0.99\textwidth]{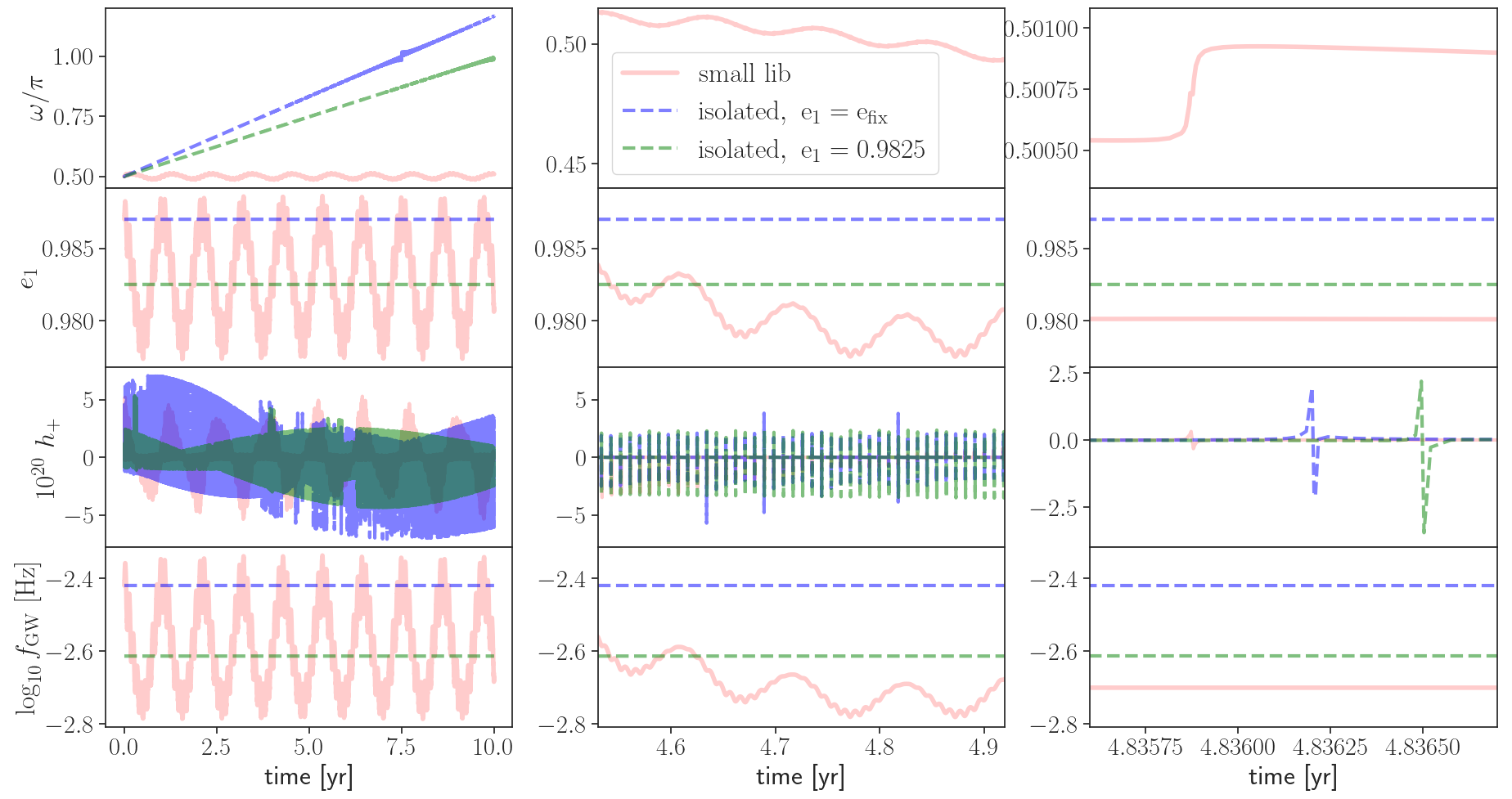}
    \caption{Comparison for the small libration orbit (red) to the isolated orbits without the SMBH with the same initial condition on $e_1$ at the fixed point (dashed blue), or with a lower eccentricity close to the average of the oscillations (dashed green).  Left to right show zoomed in insets of time. Top to bottom: The argument of pericentre $\omega$,  eccentricity $e_1$, the '+' strain polarisation and the GW frequency.}\label{fig10}
\end{figure*}

\subsection{Comparison to an isolated binary} \label{4.3}
Figure \ref{fig10} shows the evolution of the small libration orbit compared to an isolated eccentric binary (without the SMBH). We also use the isolated case as a benchmark for the strain calculation which show a smooth modulation on top of the burst at each pericentre passage. We test two initial values of the eccentricity: the same as in the small libration case (at the fixed point; dashed blue) and at a lower value closer to the mean around the secular oscillations (dashed green). We see that in the isolated cases  $\omega$ precesses due to the 1PN term, and a larger eccentricity leads to faster precession, due to the $(1-e_1^2)^{-1}$ term in the 1PN precession rate. The eccentricity in the isolated case is almost fixed due to lack of perturbations, contrary to the small oscillations in the SMBH case. Similar behaviour is evident in $f_{\rm GW}$ (bottom panels) where the GW frequency is varying mostly due to the eccentricity oscillations but remains largely fixed for the isolated binaries. The strain polarisation looks dense with characteristic timescale of modulation either with the ZLK oscillations (SMBH case) or due to the 1PN precession (isolated cases). We also note that indeed the secular timescale is much faster than the 1PN precession, which makes $\epsilon_{\rm GR}$ (Eq. \ref{eps_gr}) small, as expected.

Besides the 1PN precession timescale and secular ZLK timescale, we also observe additional timescale in the time series. First, the strain polarisation is bursty with typical time between burst being the inner period of $P_1 \approx 0.00918\ \rm yr = 3.36\ \rm days$. Small 'sawtooth' oscillations in the eccentricity and $f_{\rm GW}$ are evident on this timescale. The eccentricity also oscillates on a longer timescale of $P_2 \approx 0.23\ \rm yr$. At every pericentre, the orbital elements change rapidly and the strain increases by several orders of magnitude for a short duration of $\sim r_p/v_p$ due to the large accelerations involved. Zooming in on one peak, we see that the arrival times (and also strain amplitudes) are well separated by about 2 hours from each other,  which is about $\sim 2.5\%$ of the total orbital period. Thus, the different orbits can be well separated from each other.


\section{Summary and Conclusions} \label{sec5}

We have explored the complex behaviour of BBHs in galactic nuclei influenced primarily by the tertiary SMBH, in the process adding to the \texttt{TSUNAMI} code the capability to produce strain signals. Our main conclusions are as follows:

$\bullet$ Even for hierarchical systems, the secular approximation can yield incorrect results due to short-term oscillations \citep[e.g.,][]{luo16,gri18,man22} especially when the BBH is in the highly eccentric phase. 

$\bullet$ In addition to the standard von Zeipel-Lidov-Kozai (ZLK) evolution where the inner argument of pericentre $\omega$ is circulating (Figure \ref{circ}), $\omega$ and the eccentricity $e_1$ can also librate with large (Figure \ref{large_lib}) or small (Figure \ref{smal_lib}) amplitude around a fixed point in phase space, which is given by Eq. (\ref{efix}) and $\omega=\pi/2,3\pi/2$ \cite[see also][]{gri2024a}. Moreover, for extremely high eccentricity (or low $j_z$), GW emission at pericentre can lead to GW capture and merger even an initially librating orbit (Figure \ref{merge}), which could potentially be observed both by LISA and ground-based GW detectors. These four families have unique signatures in $\omega$ and GW strain (Figure \ref{all_omega}).

$\bullet$ We added a method to calculate the strain polarisations within the N-body code \texttt{TSUNAMI} \citep{tsunami-code} by using the quadrupole formula and accounting for the direct accelerations from the tertiary and due to post-Newtonian effects up to 3.5PN. Using this method we are able to compare the different GW signatures directly from the simulations (Figure \ref{fig8}). We find that most of the strain signatures include high bursts separated by a quiescent phase on timescales of years. The small-amplitude orbit maintains its high eccentricity and has consistent burst on the order of its inner orbital period of a few days.

$\bullet$ Due to the complexity of the dynamics, each orbit has almost unique signature. Varying the SMBH mass by as little as $3\%$ ($\gamma=0.99$ in Figure \ref{fig9}) will lead to an offset of the  time-of-arrival of the burst by about $20$ s,  which can still enable distinguishability, based on the precision to which we are able to isolate the arrival time of a single burst. Distinguishing BBH in galactic nuclei from those formed in more equal-mass field triples is also possible due to the breakdown of the test-particle limit and the backreaction on the tertiary orbit (Figure \ref{fig9}). Moreover, the orbits of BBHs with a SMBH perturber are markedly different from a binary in isolation with the same initial conditions (Figure \ref{fig10}). 

$\bullet$ The GW signature should be distinguishable in the LISA band where the SNR is high enough (around 100 for peak frequency, Figure \ref{char}). Semi-analytical population studies suggest that $10\%$ of BBHs in the Galactic Centre (or 1000 BBHs in total) will have highly eccentric librating orbits (Figure \ref{pop_zlk}).

The duration of a burst is about $r_p/v_p$ at periapse, which is $\approx1$~minute, as in Figure \ref{fig9}. This occurs every $\sim 3.4$ days, or $>4800$ minutes. Generating 1 year of data takes $\approx0.5$ seconds for output snapshots above $\Delta t = 5,000$ sec, and linearly increases to $\sim 4$ minutes for the desired output of $\Delta t = 10\ \rm s $ in order to capture the burst time derivative. Thus, full Bayesian parameter estimation is not feasible using the waveform model we have developed. However, searches and inference methods that rely on bursts and burst timing, rather than generation of full waveforms, can identify these sources and potentially infer the properties of the system \citep{rs2023, knee2024, saini25}.


\section*{Acknowledgements}

We thank Roi Basha for pointing out the non-trivial analogy between the ZLK oscillations and the pendulum model. We thank Alan Knee, Smadar Naoz and Jess McIver for useful discussions. EG acknowledges support from the ARC Discovery Program DP240103174 (PI: Heger), the ARC OzGrav Centre of Excellence CE230100016, and the ARC Discovery Early Career Research Award (DECRA) DE260101802. IMR-S acknowledges support from the Herchel Smith Fund, the Science and Technology Facilities Council grant number ST/Y001990/1, and the Science and Technology Facilities Council Ernest Rutherford Fellowship grant number UKRI2423. AAT acknowledges support from the Horizon Europe research and innovation programs under the Marie Skłodowska-Curie grant agreement no. 101103134.
\section*{Data Availability}
The data underlying this article will be provided by the corresponding author upon reasonable request.

\appendix
\section{Gravitational-wave strain construction from the quadrupole projection}

We compute the two polarizations of the gravitational-wave strain, $h_{+}$ and $h_{\times}$, from the leading-order (mass) quadrupole formula, following the tensorial notation and transverse--traceless (TT) projection conventions of \citet{maggiore}. In the wave zone at distance $R$ from the source, the radiative metric perturbation is
\begin{equation}
	h^{\rm TT}_{ij}(t,\mathbf{n})=\frac{2G}{c^{4}R}\,\mathcal{P}_{ijkl}(\mathbf{n})\,\ddot{I}_{kl}(t)\,,
\end{equation}
where $I_{ij}$ is the (trace-free) mass quadrupole moment of the source, $\mathbf{n}$ is the unit vector pointing from the source to the observer, and $\mathcal{P}_{ijkl}$ is the TT projector,
\begin{equation}
	\mathcal{P}_{ijkl}(\mathbf{n}) \equiv P_{ik}P_{jl}-\frac{1}{2}P_{ij}P_{kl},\qquad
	P_{ij}\equiv \delta_{ij}-n_{i}n_{j}\,.
\end{equation}

At leading order, one may write $\ddot{I}_{ij}$ directly in terms of instantaneous $(\mathbf{x},\mathbf{v},\mathbf{a})$ without finite differencing in time.

Given the sky location $(\theta,\phi)$ of the observer in the source frame, we define the propagation direction
\begin{equation}
	\mathbf{n}=\big(\sin\theta\,\sin\phi,\ \sin\theta\,\cos\phi,\ \cos\theta\big)\,,
\end{equation}
and construct an orthonormal triad $(\mathbf{p},\mathbf{q},\mathbf{n})$ such that $\mathbf{p}\cdot\mathbf{n}=\mathbf{q}\cdot\mathbf{n}=0$ and $\mathbf{p}\cdot\mathbf{q}=0$. Operationally, we pick an arbitrary reference vector $\mathbf{t}$ not (nearly) parallel to $\mathbf{n}$, form
\begin{equation}
	\mathbf{p}=\frac{\mathbf{t}\times\mathbf{n}}{\lVert\mathbf{t}\times\mathbf{n}\rVert},
	\qquad
	\mathbf{q}=\frac{\mathbf{n}\times\mathbf{p}}{\lVert\mathbf{n}\times\mathbf{p}\rVert},
\end{equation}
and then define the usual plus and cross polarization tensors
\begin{equation}
	e^{+}_{ij}=p_{i}p_{j}-q_{i}q_{j},
	\qquad
	e^{\times}_{ij}=p_{i}q_{j}+q_{i}p_{j}.
\end{equation}
These satisfy $e^{A}_{ij}n^{j}=0$ and $e^{A}_{ii}=0$ and provide a basis for the TT subspace.

Finally, we construct the (already trace-free) radiative tensor
\begin{equation}
	H_{ij}\equiv \frac{2G}{c^{4}R}\,\ddot{I}_{ij},
\end{equation}
and obtain the two polarizations by contraction with the polarization tensors,
\begin{equation}
	h_{+}=\frac{1}{2}\,H_{ij}\,e^{+}_{ij},
	\qquad
	h_{\times}=\frac{1}{2}\,H_{ij}\,e^{\times}_{ij}\,.
	\label{eq:hplus_hcross}
\end{equation}



\bibliographystyle{mnras}


\bsp	
\label{lastpage}
\end{document}